\documentclass[aps,prl,twocolumn,showpacs,floatfix,superscriptaddress]{revtex4}

\usepackage{graphicx,amsmath,bbm,mathrsfs,amssymb,psfrag,stmaryrd}

\newcommand{\ket}[1]{\vert #1 \rangle}

\def\nh{\hat{n}}
\def\beq{\begin{equation}}
\def\eeq{\end{equation}}
\def\beqq{\begin{eqnarray}}
\def\eeqq{\end{eqnarray}}
\begin{document}

\title{Quantum contextuality in $N$-boson systems}

\author{Fabio Benatti}
\affiliation{Dipartimento di Fisica, Universit\`a degli Studi di
Trieste, I-34151 Trieste, Italy}
\affiliation{Istituto Nazionale di Fisica Nucleare, Sezione di
  Trieste, I-34014 Trieste, Italy}

\author{Roberto Floreanini}
\affiliation{Istituto Nazionale di Fisica Nucleare, Sezione di
  Trieste, I-34014 Trieste, Italy}

\author{Marco Genovese}
\affiliation{INRIM, Strada delle Cacce 91, I-10135 Torino, Italy}

\author{Stefano Olivares}
\affiliation{Dipartimento di Fisica, Universit\`a degli Studi di
Trieste, I-34151 Trieste, Italy}


\begin{abstract}
  Quantum contextuality in systems of identical bosonic particles is
  explicitly exhibited via the maximum violation of a suitable
  inequality of Clauser-Horne-Shimony-Holt type. Unlike the approaches
  considered so far, which make use of single-particle observables,
  our analysis involves collective observables constructed using
  multi-boson operators. An exemplifying scheme to test this violation
  with a quantum optical setup is also discussed.
\end{abstract}


\maketitle

A physical theory is called noncontextual if the measurement results
about a given observable $A$ do not depend on whether other commuting
observables $B$ are simultaneously measured. Quantum Mechanics turns
out to be a contextual theory, Bell's nonlocality being a particular
form of contextuality~\cite{KS, Mermin1, Genovese}.
\par
Many experimental tests of contextuality have been proposed, where the
commuting observables $A$ and $B$ refer to the spatial and
polarization degrees of freedom of single spin-$\frac12$ particles or
photons~\cite{Basu, Benatti, Rauch, Michler, Gadway, Karimi, Home,
  Cabello1, Cabello2, Cabello3, Simon}.  Most of the tests check
whether inequalities of Clauser-Horne-Shimony-Holt (CHSH)
type~\cite{CHSH}, applied now to local, single-system settings:
\begin{equation} \label{CHSH}
\mathcal{E}=\Big|\langle A(B+B')\rangle+\langle
A'(B-B')\rangle\Big|\leq 2\ ,
\end{equation}
are violated or not, $\langle\cdots\rangle$ being the expectation
values of the involved quantities in a suitable system state. These
inequalities are a straightforward consequence of assuming that the
values assigned to products $AB$, $AB'$, $A'B$ and $A'B'$ of
observables $A$ and $A'$, pertaining to a given degree of freedom,
commuting with observables $B$ and $B'$ pertaining to a different
degree of freedom, are products of the values independently assigned
to the separate observables $A$, $A'$, $B$ and $B'$.
\par
Note that different approaches to investigate quantum contextuality,
not based on the inequality (\ref{CHSH}), have also been considered,
for instance, see
\cite{Mermin2,Blatt,Bartosik,Lim,Huang,Guhne,Plastino,Paris}. Furthermore,
violation of noncontextuality has been exploited as a resource for
quantum key distribution and quantum information processing purposes
\cite{Adhikari,Kleinmann,Horodecki}.  The contextuality of mesoscopic
systems has been recently addressed in systems consisting of
distinguishable particles~\cite{Cabello4} and superconductive
devices~\cite{Wei}, but all involving single-qubit measurements. 

The question which naturally arises is whether it is possible to test
noncontextuality in systems of $N$ identical particles, 
like trapped ultracold bosons, phonons in systems of micro-mechanical oscillators
and multiphoton states in quantum optics, {\it i.e.} in truly mesoscopic
systems. In such cases, the use of collective observables is mandatory;
from the theoretical point of view, 
these observables can be naturally described in
terms of multiboson algebras, which represent a novel tool in this
specific context. This analysis of mesoscopic systems represents
also an important contribution to the study of the transition
between quantum and classical systems.
\par
Motivated by the previous considerations, here we address
quantum contextuality in systems made of $N$ identical particles
(bosons). Our main result is the identification
of a set of collective observables for multi-particle tests of
noncontextuality.
More specifically, using maximally entangled NOON-like
states, we first show that inequalities of CHSH type as in
(\ref{CHSH}) can not be violated if constructed with standard
single-boson observables; this is ultimately due to the
indistinguishability of the involved particles.  In order to make
quantum contextuality apparent in such systems, one has to use instead
collective observables built with suitable multi-boson operators;
indeed, with such observables one can construct CHSH like inequalities
that are maximally violated by the same NOON-like states.  A physical
implementation based on a quantum optical interferometric scheme able
to test these results is also presented and discussed.
\par
Let us consider a system of $N$ bosons with two degrees of freedom,
{\it e.g.} a ``spatial'' and an ``internal'' one. For sake of
definiteness, we assume these degrees of freedom to take two possible
values, labeled by the variables $k=1, 2$ and $\alpha=-,+$,
respectively. As appropriate for boson systems, we adopt a second
quantized description and introduce suitable creation
$a_k^{(\alpha)}{}^\dagger$ (and annihilation $a_k^{(\alpha)}$)
operators for the mode $(k,\alpha)$, satisfying the usual bosonic
algebra: $\big[a_k^{(\alpha)},\
a_l^{(\beta)}{}^\dagger\big]=\delta_{k,l}\delta_{\alpha,\beta}$.  Out
of the vacuum $|0\rangle$, they create the (Fock) states of the system
\beq
\label{not1}
\Big|n_1^{(-)},n_2^{(-)},n_1^{(+)},n_2^{(+)}\Big\rangle=
\prod_{k=\{1,2\}}\prod_{\alpha=\{-,+\}}
\frac{\big(a_k^{(\alpha)}{}^\dagger\big)^{n_k^{(\alpha)}}}{\sqrt{n_k^{(\alpha)}!}}|0\rangle\
, \eeq
containing $n_k^{(\alpha)}$ bosons in the mode $(k,\alpha)$,
such that $\sum_{k,\alpha} n_k^{(\alpha)}=N$. For sake of compactness,
hereafter we shall indicate only the nonvanishing occupation numbers,
{\it i.e.}, those for which $n_k^{(\alpha)}\neq 0$.
\par
A basic example of a quantum system exhibiting contextuality, that is
violating the expectation that the result of a measurement be
independent of a previous or simultaneous measurement of any set of
mutually commuting observables, is a single particle system ($N=1$),
prepared in the following superposition of Fock states:
\begin{align}
\ket{\psi}&=\frac{1}{\sqrt{2}} \Big(a_1^{(-)}{}^\dagger + a_2^{(+)}{}^\dagger\Big)\ket{0} \nonumber\\
&=\frac{1}{\sqrt{2}} \Big(\ket{n_1^{(-)}=1}+\ket{n_2^{(+)}=1}\Big)\ .
\label{psi:1}
\end{align}
By taking the variables $k$ and $\alpha$ to label suitable spin-like
components, one finds that CHSH-like inequalities as in (\ref{CHSH})
are indeed maximally violated, as confirmed in experiments based on
single-neutron \cite{Rauch} and single-photon interferometers
\cite{Michler, Gadway}.
\par
Similarly, in the case of $N$ bosons one can consider the following
spin-like observables
\begin{subequations}
\label{svar}
\begin{align}
J_x &= \frac12 \sum_{\alpha=\{-,+\}} \Big[a_1^{(\alpha)}{}^\dagger\, a_2^{(\alpha)} +
a_2^{(\alpha)}{}^\dagger\, a_1^{(\alpha)}\Big]\ ,\\
J_y &= \frac i2 \sum_{\alpha=\{-,+\}} \Big[a_2^{(\alpha)}{}^\dagger\, a_1^{(\alpha)} -
a_2^{(\alpha)}{}^\dagger\, a_1^{(\alpha)}\Big]\ ,\\
J_{z} &= \frac12 \sum_{\alpha=\{-,+\}} \Big[a_1^{(\alpha)}{}^\dagger\, a_1^{(\alpha)} -
a_2^{(\alpha)}{}^\dagger\, a_2^{(\alpha)}\Big]\ ,\\
S_{x} &= \frac12 \sum_{k=\{1,2\}} \Big[a_k^{(-)}{}^\dagger\, a_k^{(+)} +
a_k^{(+)}{}^\dagger\, a_k^{(-)}\Big]\ ,\\
S_{y} &= \frac i2 \sum_{k=\{1,2\}} \Big[a_k^{(+)}{}^\dagger\, a_k^{(-)} -
a_k^{(-)}{}^\dagger\, a_k^{(+)}\Big]\ ,\\
S_{z} &= \frac12 \sum_{k=\{1,2\}} \Big[a_k^{(-)}{}^\dagger\, a_k^{(-)} -
a_k^{(+)}{}^\dagger\, a_k^{(+)}\Big]\ .
\end{align}
\end{subequations}
The observables $J_i$ and $S_i$ both satisfy the $SU(2)$ algebraic
relations ($[J_i,J_j]=i\epsilon_{ijk}J_k$, and similarly for $S_i$),
while commuting among themselves, $[J_i,S_j]=\,0$.  Furthermore, their
eigenvalues are the integers between $-N/2$ and $N/2$. Then, once
rescaled by $2/N$, their eigenvalues lie in the interval $[-1,1]$ and
the rescaled observables should violate a suitable constructed
inequality of CHSH type of the form (\ref{CHSH}).
\par
In the case of the single-boson system ($N=1$), this can be easily
shown by taking the following choices for the four observables
appearing in the inequality (\ref{CHSH}):
\begin{align*}
&A = 2\, J_z\ ,\phantom{'}\quad B=2\, S_z(\pi/4)\ ,\\
&A'= 2\, J_x\ ,\quad B' = 2\, S_z(-\pi/4)\ ,
\end{align*}
where $S_{z}(\theta) = \cos\theta\, S_{z} + \sin\theta\, S_{x}$.
For the state (\ref{psi:1}), straightforward calculations lead to:
\begin{align}
\langle J_{z} \otimes S_{z} \rangle &=
\langle J_{x} \otimes S_{x} \rangle = 1/4\ , \\
\langle J_{z} \otimes S_{x} \rangle &=
\langle J_{x} \otimes S_{z} \rangle = 0\ ,
\end{align}
whence ${\cal E} = 2\sqrt{2} > 2$ indeed exhibits the maximum
violation of the CHSH inequality allowed by quantum
mechanics~\cite{Cirelson}.
\par
For systems of bosons with $N>1$, the obvious generalization of these
single-particle choices does not lead to any noncontextuality
violation; this is not surprising, since in order to exhibit quantum
contextuality the CHSH test, ${\cal E}>2$, needs to be adapted to the
system under study through a careful choice of both the state and of
the four observables $A$, $A'$, $B$ and $B'$.  In the case of $N$
bosons, the state (\ref{psi:1}) can be easily generalized:
\begin{align}
\ket{\Psi}&=\frac{1}{\sqrt{2N!}} \Big[ \big(a_1^{(-)}{}^\dagger\big)^N 
+ \big(a_2^{(+)}{}^\dagger\big)^N\Big]\ket{0} \nonumber\\
&=\frac{1}{\sqrt{2}} \Big[\ket{n_1^{(-)}=N}+\ket{n_2^{(+)}=N}\Big]\ ;
\label{NOON}
\end{align}
this is a NOON-like state, superposition of two Fock states, the first
representing $N$ bosons all in the same mode with $k=1$, $\alpha=-$
and similarly the other, $N$ bosons in the state with $k=2$,
$\alpha=+$; it clearly reduces to the state (\ref{psi:1}) when
$N=1$. However, the averages of the same observables as before in this
new state $|\Psi\rangle$ become:
\begin{align}
\langle J_{z} \otimes S_{z} \rangle &= N^2/4\ ,\\
\langle J_{x} \otimes S_{x} \rangle &=
\langle J_{z} \otimes S_{x} \rangle =
\langle J_{x} \otimes S_{z} \rangle = 0\ ;
\end{align}
as a consequence, after the necessary rescaling of the spin-like
observables by $2/N$, one finds ${\cal E} = \sqrt{2} < 2$ and
therefore no violation of the noncontextuality test.
\par
Consider instead the multi-boson operators~\cite{Brandt, Rasetti1,
  Rasetti2}: \beq
\label{collop0}
A_k^{(\alpha)} = F_N\big(\nh_k^{(\alpha)}\big)\ \big(a_k^{(\alpha)}\big)^N\ ,
\eeq
where 
\begin{equation}
\label{collop1}
F_N\big(\nh_k^{(\alpha)}\big) = \Bigg[
\bigg\llbracket{\frac{\nh_k^{(\alpha)}+N}{N}}\bigg\rrbracket
\frac{\nh_k^{(\alpha)}!}{(\nh_k^{(\alpha)}+N)!}
\Bigg]^{1/2},
\end{equation}
with $\nh_k^{(\alpha)}=a_k^{(\alpha)}{}^\dagger\, a_k^{(\alpha)}$ the
number operator relative to the mode $(k, \alpha)$, the symbol
$\llbracket{\ }\rrbracket$ representing the integer part.  The
$A_k^{(\alpha)}$ are collective annihilation operators as they
diminish by $N$ the number of bosons; indeed, on the orthonormal basis
of Fock number states as in~(\ref{not1}), they act as follows: \beq
\label{collop2}
A_k^{(\alpha)}\,\big|n_k^{(\alpha)}\big\rangle=\sqrt{\bigg\llbracket{\frac{n_k^{(\alpha)}}{N}}\bigg\rrbracket}\ 
\big|n_k^{(\alpha)}-N\big\rangle\ ,
\eeq
while the Hermitian conjugate operators $A_k^{(\alpha)}{}^\dagger$ act as collective creation operators:
\beq
\label{collop2b}
A_k^{(\alpha)}{}^\dagger\,\big|n_k^{(\alpha)}\big\rangle=
\sqrt{\bigg\llbracket{\frac{n_k^{(\alpha)}}{N}}\bigg\rrbracket +1}\ 
\big|n_k^{(\alpha)}+N\big\rangle\ .
\eeq
Then, one easily finds that:
$\Big[A_k^{(\alpha)},A_l^{(\beta)}{}^\dagger\Big] =
\delta_{k,l}\,\delta_{\alpha,\beta}$. 
\par
In analogy with (\ref{svar}), with these multi-boson annihilation and creation operators
one can now form collective spin-like operators relative to the spatial degree of freedom:
\begin{subequations}
\label{J:q}
\begin{align}
{\cal J}_x &= \frac12 \sum_{\alpha=\{-,+\}} \Big[A_1^{(\alpha)}{}^\dagger\, A_2^{(\alpha)} +
A_2^{(\alpha)}{}^\dagger\, A_1^{(\alpha)}\Big]\ ,\\
{\cal J}_y &= \frac i2 \sum_{\alpha=\{-,+\}} \Big[A_2^{(\alpha)}{}^\dagger\, A_1^{(\alpha)} -
A_2^{(\alpha)}{}^\dagger\, A_1^{(\alpha)}\Big]\ ,\\
{\cal J}_{z} &= \frac12 \sum_{\alpha=\{-,+\}} \Big[A_1^{(\alpha)}{}^\dagger\, A_1^{(\alpha)} -
A_2^{(\alpha)}{}^\dagger\, A_2^{(\alpha)}\Big]\ ;
\end{align}
\end{subequations}
one checks that they satisfy the $SU(2)$ commutation relations, 
$[{\cal J}_i, {\cal J}_j] = i\epsilon_{ijk} \,{\cal J}_k$. 
In the same way, the multi-boson, collective spin-like polarization operators
\begin{subequations}
\label{S:q}
\begin{align}
{\cal S}_{x} &= \frac12 \sum_{k=\{1,2\}} \Big[A_k^{(-)}{}^\dagger\, A_k^{(+)} +
A_k^{(+)}{}^\dagger\, A_k^{(-)}\Big]\ ,\\
{\cal S}_{y} &= \frac i2 \sum_{k=\{1,2\}} \Big[A_k^{(+)}{}^\dagger\, A_k^{(-)} -
A_k^{(-)}{}^\dagger\, A_k^{(+)}\Big]\ ,\\
{\cal S}_{z} &= \frac12 \sum_{k=\{1,2\}} \Big[A_k^{(-)}{}^\dagger\, A_k^{(-)} -
A_k^{(+)}{}^\dagger\, A_k^{(+)}\Big]\ .
\end{align}
\end{subequations}
satisfy $[{\cal S}_i, {\cal S}_j] = i\epsilon_{ijk} \,{\cal S}_k$. 
Further, as before, the two sets
of operators mutually commute, $[{\cal J}_i,\, {\cal S}_j]=\, 0$.
\par
Further, notice that
\beq
\label{numberop}
A_k^{(\alpha)}{}^\dagger\, A_k^{(\alpha)}=\bigg\llbracket{\frac{\nh_k^{(\alpha)}}{N}}\bigg\rrbracket\ , 
\eeq 
so that, unlike the single-boson spin-like
operators in (\ref{svar}), the operators in~(\ref{J:q}) and~(\ref{S:q})
have eigenvalues $\pm1/2$.
Thus, we can directly consider the inequality~(\ref{CHSH}) relative to the
NOON state (\ref{NOON}) and to the collective observables
\begin{subequations}
\label{CHSH_N}
\begin{align}
&A = 2\, {\cal J}_z\ ,\phantom{'}\quad B=2\, {\cal S}_z(\pi/4)\ ,\\
&A'= 2\, {\cal J}_x\ ,\quad B' = 2\, {\cal S}_z(-\pi/4)\ ,
\end{align}
\end{subequations}
where ${\cal S}_z(\theta) = \cos\theta\, {\cal S}_z + \sin\theta\, {\cal S}_x$.
One finds:
\begin{align}
\langle {\cal J}_z \otimes {\cal S}_z \rangle &=
\langle {\cal J}_x \otimes {\cal S}_x \rangle = 1/4 \\
\langle {\cal J}_z \otimes {\cal S}_x \rangle &=
\langle {\cal J}_x \otimes {\cal S}_z \rangle = 0 \ ,
\end{align}
whence ${\cal E}= 2\sqrt{2} > 2$, so that the
CHSH inequality (\ref{CHSH}) results maximally violated
and quantum contextuality is manifest.
\par
An example of physical setup able to test, in principle, this
violation can be built within quantum optics. In this framework, the
two degrees of freedom labeled by $k$ and $\alpha$ can be taken to
refer to the photon path (momentum) and polarization: the operator
$a_k^{(\alpha)}{}^\dagger$ creates from the vacuum a single photon
along the path $k$ with polarization $\alpha$, while
$A_k^{(\alpha)}{}^\dagger$ creates $N$ photons in the same
state. Similarly, unlike $J_i$ and $S_i$ in (\ref{svar}), the
operators ${\cal J}_i$ in (\ref{J:q}) and ${\cal S}_i$ in (\ref{S:q})
are collective observables, referring to path and polarization degrees
of freedom, respectively.
\par
Notice that the experimental implementation
of the test does not require the actual realization of the
multi-boson operators $A_k^{(\alpha)}{}^\dagger$ and $A_k^{(\alpha)}$, 
but only of suitable procedures for measuring the collective observables
${\cal J}_i$ and ${\cal S}_i$. In order to achieve this,
operations acting on the $N$-photon states as a whole are nevertheless needed; 
as a consequence, linear optical passive devices cannot directly be used.
\par
\begin{figure}
\includegraphics[width=8cm]{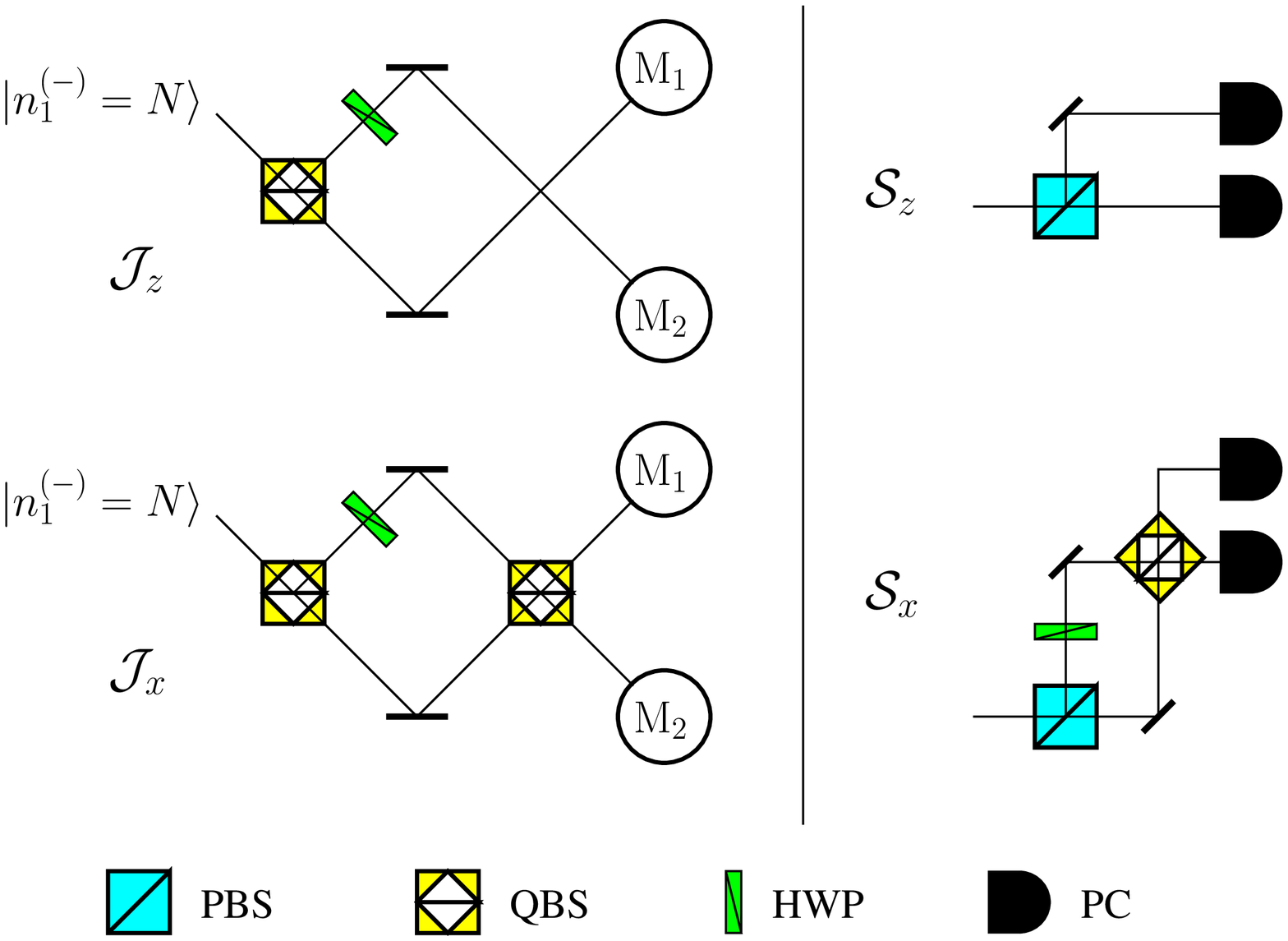}
\caption{\label{f:scheme} Physical implementation of a quantum optics
  setup able to test quantum contextuality with NOON-like
  states. Legend of the components: PBS~=~Polarizing Beam Splitter,
  QBS~=~Quantum Beam Splitter, HWP~=~Half-Wave Plate, PC~=~Photon
  Counter. See the text for details.}
\end{figure}
\par
The proposed implementation, sketched in
Fig.\ref{f:scheme}, is based on a Mach-Zehnder-like interferometric
scheme; however, the usual beam splitters are replaced with the ``quantum
beam splitters'' (QBSs) proposed in \cite{Dunningham}, where
a nonlinear medium is inserted in one of the two arms. 
These quantum devices, realized through the unitary operator $U_{\rm QBS}$, 
generates NOON-like states by acting on a Fock
state where the $N$ photons are all prepared in a given 
path-polarization configuration $(k,\alpha)$ \cite{Dunningham}:
\begin{align}
\label{qbs-state}
U_{\rm QBS}\, \big|n_k^{(\alpha)}=N\big\rangle =
\frac{\big|n_k^{(\alpha)}=N\big\rangle+(-1)^l\big|n_l^{(\alpha)}=N\big\rangle}{\sqrt{2}},
\end{align}
with $l\ne k$ and $k,l=1,2$. Here, $k=1$ ($k=2$) refers to the lower
(upper) path, while $\alpha=-,+$ to the two possible 
horizontal and vertical polarization,
{\it i.e.} $H\equiv -$, $V\equiv +$.
The NOON-like state given in (\ref{NOON}) can then be obtained from 
the appropriate state (\ref{qbs-state}) with $k=1$ and $\alpha=-$, by switching
the polarization in the upper ($k=2$) path 
by means of a half-wave plate (HWP).
[It is worth noticing that, due to the presence of the
two mirrors which exchange the modes $1$
and $2$ (see Fig.\ref{f:scheme}), the lower (upper) path is detected as $k=2$ ($k=1$); in order
to avoid confusion, in the Fig.~\ref{f:scheme} we call the detectors
$M_k$, with $k$ referring to the actual followed path.]
\par
The operator ${\cal J}_z$ can then be measured by sending the outgoing beams
directly to the two detector devices $M_k$ (upper-left scheme in Fig.\ref{f:scheme});
indeed, the detection operation at $M_1$ and $M_2$, for instance implemented by photon counters,
precisely corresponds to the projection onto the eigenstates of ${\cal J}_z$.
Instead, in the case of the observable ${\cal J}_x$
one needs to make the two beams interfere at
another QBS (lower-left scheme in Fig.\ref{f:scheme}) before detection at $M_k$.
Using (\ref{qbs-state}), the outgoing state after the second QBS reads now 
\begin{equation}
\ket{\Psi_{\rm out}} =
\frac{1}{\sqrt{2}}
\left(\big|\Psi_2^{(-)}\big\rangle - \big|\Psi_1^{(+)}\big\rangle\right),
\end{equation}
where $\big|\Psi_1^{(+)}\big\rangle = 
\big(\big|n_1^{(+)}=N\big\rangle - \big|n_2^{(+)}=N\big\rangle\big)/\sqrt{2}$ and
$\big|\Psi_2^{(-)}\big\rangle = 
\big(\big|n_1^{(-)}=N\big\rangle + \big|n_2^{(-)}=N\big\rangle\big)/\sqrt{2}$
are eigenstates of ${\cal J}_x$, the label $k$ in $\big|\Psi_k^{(\pm)}\big\rangle$, $k=1,2$,
referring now to the upper and lower exiting paths, beyond the additional QBS.
As a result, the detection operation at $M_1$ and $M_2$
corresponds to projections onto these two states.
\par
In order to simultaneously determine, together with ${\cal J}_z$,
$i=x,z$, also the operators ${\cal S}_z$ and ${\cal S}_x$, as required
by the inequality (\ref{CHSH}), the detection procedures at $M_k$ need
to be more structured than those provided by simple photon counters.
The measurement of ${\cal S}_z$ can be implemented through a
polarizing beam splitter (PBS) and two photon counters (PCs), that
allows to discriminate between $\big|n_k^{(-)}=N\big\rangle$ and
$\big|n_k^{(+)}=N\big\rangle$, eigenstates of ${\cal S}_z$ (see the
right-upper scheme in Fig.\ref{f:scheme}).  Instead, in order to
measure ${\cal S}_x$, one needs to insert before the photon counters a
further Mach-Zehnder interferometric circuit, built with additional
HWP and QBS (see the right-lower scheme of the Fig.~\ref{f:scheme}).
From the same considerations described before, in explaining the
determination of ${\cal J}_x$, now applied to the polarization degrees
of freedom, one can check that photon detection at the exit arms of
this additional interferometer precisely corresponds to projection
onto the eigenstates of ${\cal S}_x$,
$\big(\big|n_k^{(-)}=N\big\rangle \pm
\big|n_k^{(+)}=N\big\rangle\big)/\sqrt{2}$.
\par
Finally, it is worth noticing that the schemes just illustrated for
measuring the relevant ${\cal J}_i$ and ${\cal S}_i$ observables are
such that all the $N$ photons arrive at one and only one of the photon
counters. Therefore, by suitably composing one of the two detecting
schemes in the right panel of the Fig.~\ref{f:scheme} with one of the
measurements of the left panel, one can experimentally determine all
collective observables needed to test the CHSH inequality discussed
above.
\par
Summarizing, in this Letter we discussed a test of quantum
contextuality in systems composed by $N$ identical bosonic
particles. We showed that, unlike the analysis so far considered in
the literature, based on systems of distinguishable particles using
single-particle observables, in order to make quantum contextuality
apparent in such mesoscopic bosonic systems, measurements of
collective observables are needed. We further remark that the
observables we used are built out of collective operators belonging to a
multiboson algebra, a tool that has never been considered in such a
context. Incidentally, a state-independent test of
noncontextuality with $N$-boson
systems can also be straightforwardly achieved with our
collective observables following, {\it e.g.}
Refs.~\cite{Cabello1,Cabello3}.

We are confident that these theoretical results will open
new perspectives in the study of quantum contextuality in mesoscopic
systems, as trapped ultracold bosons and phonons in systems 
of micro-mechanical oscillators, and provide new insights
in the actual realization of further experimental tests, possibly paving the
way to new applications in quantum technology.

\end{document}